\documentclass[final,1p,times]{elsarticle} 

\newcommand{\sqrtsNN}{\mbox{$\sqrt{\mathrm{s}_{_{\mathrm{NN}}}}$}}
\newcommand{\axi}{$\overline{\Xi}^+$}
\newcommand{\xim}{$\Xi^-$}
\newcommand{\alam}{$\overline{\Lambda}$}
\newcommand{\lam}{$\Lambda$}
\newcommand{\ks}{$\mathrm{K}^{0}_{\mathrm S}$}

\newcommand{\ppt}{$p_{\rm T}$}

\def \pp    {$p + p$ }

\usepackage{graphicx} 
\usepackage{amssymb} 
\usepackage{amsthm} 
\usepackage{lineno} 

\journal{Nuclear Physics A} 
\begin{document} 

\begin{frontmatter} 


\title{Strangeness production in d+Au collisions at \sqrtsNN\ =\ 200 GeV in STAR}

\author{Xianglei Zhu$^{a}$ for the STAR collaboration}

\address[a]{Department of Engineering Physics, Tsinghua University, Beijing 100084, China}

\begin{abstract} 
We report on the measurements of strange hadron (\ks, $\Lambda$,
$\Xi$) production in the most central (0-20\%) d+Au collisions at
\sqrtsNN\ =\ 200 GeV in STAR. Significant strangeness enhancement is
observed in the most central d+Au collisions, especially for
multi-strange hyperons ($\Xi$). The nuclear modification factor
($R_{dAu}$) of identified particles shows possible particle type --
baryon (protons, $\Lambda$), meson ($\pi$,\ks\ ,$\phi$) --
dependence at intermediate \ppt\ (2 - 3.5 GeV/c), indicating that
the final state may have an impact on the Cronin effect.
\end{abstract} 

\end{frontmatter} 



\section{Introduction}

Enhancement of strange hadrons' yield in high energy nuclear
collisions relative to that in \pp\ collisions at the same energy
has long been regarded as one of the signatures of Quark-Gluon
Plasma (QGP) production in these collisions \cite{raf82}.
Supposedly, it will be more efficient to create $s\bar{s}$ pair in
the QGP than in hadronic matter. In experiments, substantial
strangeness enhancement has been observed in Au+Au and Cu+Cu
collisions in STAR, especially for multi-strange hyperons
\cite{Abelev:2007xp,Adams:2006ke,ant}. According to canonical
statistical model, strangeness enhancement in nuclear collisions can
also be due to canonical suppression of strangeness production in
\pp\ collisions \cite{Redlich:2001kb}. Recently, STAR's $\phi$ meson
 data \cite{Xu:2008zzd,Abelev:2008zk} shows an above unity enhancement in heavy ion collisions
at RHIC, which seems favor the approach of a fully equilibrated core
plus hadronic corona \cite{Becattini:2008ya}. The study of
strangeness production in d+Au collisions will connect A+A and \pp\
data, thus leading to better understanding of strangeness enhancement
in nuclear collisions.

The Cronin effect, which is the enhancement of hadron spectra at
intermediate \ppt\ in p+A collisions with increasing nuclear size,
was first seen in the nuclear modification factor in low energy p+A
collisions \cite{cronin}. At RHIC energy, the nuclear modification
factor in d+Au collisions for pions and protons \cite{Adams:2003qm}
show that the Cronin effect exists for both particle species. The
conventional models of the Cronin effect attribute the change of
\ppt\ spectra in p+A to multiple parton scattering in the initial stage
before hard parton scattering, which will broaden the outgoing
parton/hadron \ppt\ distribution \cite{Accardi:2002ik}. Recently,
the coalescence/recombination model, which describes the
hadronization of partons in deconfined matter and reproduces well
many features of hadron spectra, $v_2$ and $R_{CP}$ in Au+Au
collisions at RHIC, has been applied to d+Au collisions to describe
the hadron spectra and the Cronin effect \cite{Hwa:2004zd}.
Moreover, the recombination model predicts particle type
(baryon/meson) dependence of the Cronin effect, as the particle type
dependence of $R_{CP}$ or $v_2$ at intermediate \ppt\ in Au+Au
collisions at RHIC. Therefore, the measurements of the Cronin effect
for more particle species are helpful to understand the mechanism of
the Cronin effect.

In STAR, strange hadrons are reconstructed with the topology of
their weak decay channels, $K^0_s \rightarrow \pi^+\pi^-$ (69.2\%
branching ratio), $\Lambda\rightarrow p\pi$ (63.9\% branching ratio)
and $\Xi\rightarrow\Lambda\pi$ (99.9\% branching
ratio)\cite{Adler:2002uv,Adams:2003fy}. This analysis is based on
about 32 million minimum bias d+Au collision events from STAR's 2008
run. For the time being, only the most central events (0-20\%) are
well defined ($N_{part} = 15.2$ and $N_{bin} = 14.6$ according to
Glaubler model). In this paper, we present the results for the most
central 20\% events only.

\section{Results}

\begin{figure} [h]
\centerline{\includegraphics[width=0.50\textwidth]{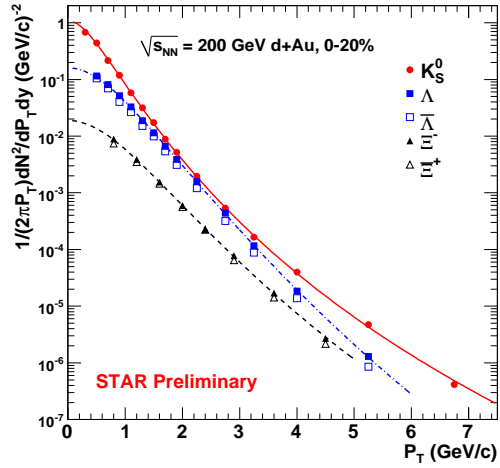}}
\caption{Transverse momentum spectra of \ks, \lam(\alam), \xim(\axi)
at $|y|<1.0$ from the most central (0-20\%) d+Au collisions at \sqrtsNN
= 200 GeV. The \lam(\alam) spectra were corrected by the week decay
from $\Xi$ and $\Xi^0$. The curves are Levy function fits to the
spectra (solid, dot-dashed and dashed are for \ks, \lam\ and \xim\
respectively). Errors are statistical only.} \label{fig1}
\end{figure}

After correcting the raw spectra by the acceptence, reconstruction
and vertex finding efficiencies, which is estimated with an embedding
method, we get the \ppt\ spectra of \ks, \lam(\alam), \xim(\axi)
shown in Fig. \ref{fig1} at mid-rapidity ($|y|<1.0$) for the most
central d+Au collisions (0-20\%). The \lam(\alam) spectra have been
corrected by the feed-down contributions from multi-strange baryon
($\Xi$ and $\Xi^0$) weak decays. We assume $\Xi^0$ make the same
contribution to $\Lambda$ as the $\Xi$'s. The contribution from
$\Omega$ is neglected in this analysis. The total feed-down
contribution from $\Xi$'s is on the order of 23\% to the inclusive
$\Lambda$ yield. The spectra can be well fitted with a Levy function
\cite{Wilk:1999dr}, as shown in Fig. \ref{fig1},
\begin{equation}
\frac{1}{2\pi
p_{T}}\frac{dN^2}{dp_{T}dy}=dN/dy\frac{(n-1)(n-2)}{2\pi
nT(nT+m(n-2))}(1+\frac{\sqrt{p^2_{T}+m^2}-m}{nT})^{-n}.
\end{equation}
In the Levy function, the soft component at lower \ppt\ in the spectra
is controlled by the parameter T, and the power law component at
higher \ppt\ is controlled by the index n. The integrated yield
($dN/dy$), dominated by the low $p_T$ production, can be extracted
from the fit. The percentage of dN/dy from extrapolation at low
$p_T$ is 10\% for \ks\ , 18\% for $\Lambda$ and 29\% for $\Xi$.

\begin{figure} [h]
\centerline{\includegraphics[width=0.7\textwidth]{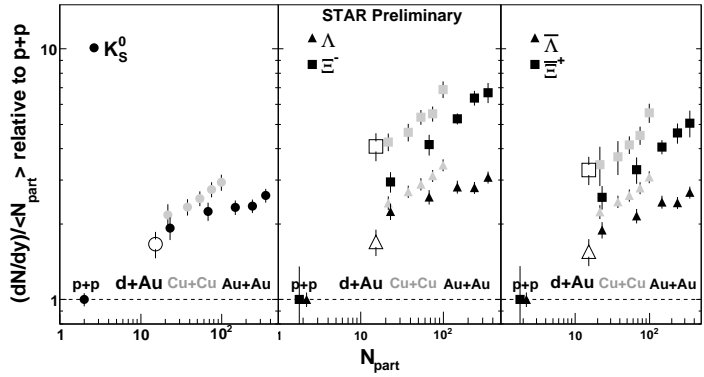}}
\caption{Strangeness enhancement factor ($dN/dy/N_{part}$ relative
to that in \pp\ collisions) for \ks\ (left), \lam\ , \xim\ (middle)
and \alam\ , \axi\ (right) from d+Au (open symbols), Cu+Cu (grey)
and Au+Au (black) collisions at \sqrtsNN = 200 GeV. For d+Au data,
errors are statistical only. The details of the Cu+Cu and Au+Au data
can be found in \cite{ant}} \label{fig2}
\end{figure}

Fig. \ref{fig2} shows the enhancement factor ($dN/dy/N_{part}$
relative to that in \pp\ collisions) as a function of $N_{part}$ for
\ks\ (left), \lam\ , \xim\ (middle) and \alam\ , \axi\ (right) from
d+Au (open symbols), Cu+Cu (grey) and Au+Au (black) collisions at
\sqrtsNN = 200 GeV. The details and interpretation of the Cu+Cu and
Au+Au data can be found in \cite{ant}. Fig. \ref{fig2} shows that
there is also significant strangeness enhancement in the most central
d+Au collisions, especially for multi-strange hyperon $\Xi$. The
enhancement factor is almost the same for \ks\ and $\Lambda$, and
much larger for $\Xi$, which shows that the enhancement factor is
proportional to the strangeness content. This strangeness content
dependence of the enhancement factor is consistent with the
predictions from the canonical statistical model
\cite{Redlich:2001kb}.

\begin{figure} [h]
\begin{center}
\includegraphics[width=9cm]{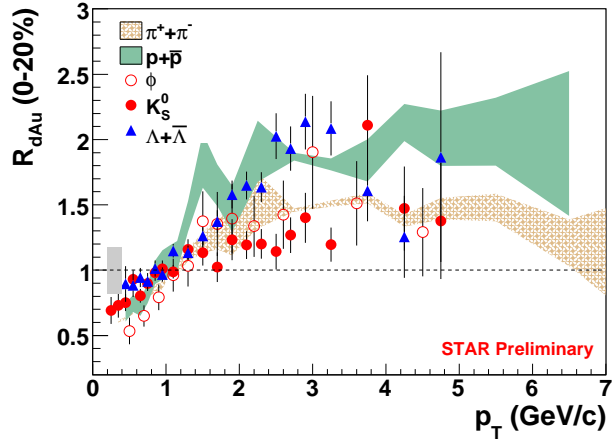}
\caption{ $R_{dAu}$ of strange hadrons (\ks,
$\Lambda+\overline{\Lambda}$, $\phi$) at mid-rapidiy from \sqrtsNN =
200 GeV d+Au collisions. $R_{dAu}$ of protons and pions are shown as
 cyan and meshed brown bands respectively \cite{Adams:2003qm}. STAR $\phi$
data is from \cite{starphi}. Errors are statistical only for pions,
protons, \ks\ and $\Lambda$. The left grey band is the 18\%
normalization error. } \label{fig3}
\end{center}
\end{figure}

Fig. \ref{fig3} shows the nuclear modification factor ($R_{dAu}$) of
strange hadrons (\ks, $\Lambda+\overline{\Lambda}$, $\phi$), as well
as those of protons and pions which are shown as cyan and meshed brown
bands respectively. STAR $\phi$ data is from \cite{starphi}. The
$R_{dAu}$ for each identified particle is similar and lower than
unity in the low \ppt\ region. When $p_{\rm T}>1$ GeV/c, the
$R_{dAu}$ values of all particles are above unity, while baryons
($\Lambda$ and proton) $R_{dAu}$ rises faster than mesons'. At
intermediate \ppt\ ($2-3.5$ GeV/c), $R_{dAu}$ is grouped into two
bands: baryons at about 1.81 $\pm$ 0.05 and mesons around 1.28 $\pm$
0.05. It should be noted here that $\phi$ mass is close to $\Lambda$
mass, but its $R_{dAu}$ is closer to \ks's. This indicates the
possible particle type dependence of $R_{dAu}$ instead of particle
mass dependence. This possible particle type (baryon/meson)
dependence of $R_{dAu}$ at intermediate \ppt\ is similar to $R_{CP}$
measurements in Au+Au collisions at the same energy
\cite{Adams:2005dq}. When \ppt\ increases beyond 3.5 GeV/c ($\sim$ 3
GeV/c for $\phi$ meson), the errors in strange hadrons' $R_{dAu}$
become too large to draw any conclusion.

\section{Summary}

In summary, we presented the transverse momentum spectra,
strangeness enhancement factor and nuclear modification factor
$R_{dAu}$ for \ks\ ,$\Lambda$ and $\Xi$ from the most central \sqrtsNN =
200 GeV d+Au collisions at RHIC. Data shows significant strangeness
enhancement in the most central d+Au collisions, especially for
multi-strange hyperon $\Xi$. The enhancement factor is almost the
same for \ks\ and $\Lambda$ and much higher for $\Xi$, which seems
consistent with the predictions from the canonical statistical model.
The $R_{dAu}$ data shows Cronin enhancement at intermediate \ppt\
and has possible particle type (baryon/meson) dependence. This
particular behavior can be naturally explained in the
coalescence/recombination model, which shows that the hadronization
process in the final state may have an impact on the Cronin effect.


\section*{Acknowledgments}
This work was in part supported by China Postdoctoral Science
Foundation.

\end{document}